\documentclass[12pt]{iopart}

\usepackage{iopams}  
\usepackage{graphicx}
\usepackage{array}
\eqnobysec
\newcommand{\noun}[1]{\textsc{#1}}

\providecommand{\tabularnewline}{\\}
\begin{document}

\title[Dynamical quintessence fields PS mass function]{Dynamical quintessence fields Press-Schechter mass function: detectability
and effect on dark haloes}

\author{Morgan Le Delliou}

\address{CFTC, Complexo Interdisciplinar, Univ.
de Lisboa, Av. Prof. Gama Pinto, 2, 1649-003 Lisboa Codex PORTUGAL}
\ead{delliou@cii.fc.ul.pt, delliou@astro.queensu.ca}
\begin{abstract}
We present an investigations on the influence of dynamical quintessence
field models on the formation of non-linear structures. In particular,
we focus on the structure traced by the mass function. Our contribution
builds on previous studies by considering two potentials not treated
\textbf{previously} and investigating the impact of the free parameters
of two other models. Our approach emphasises the physical insight
into the key role of the evolution of the equation of state%
\textbf{: we use the variations in the spherical non-linear collapse caused by
quintessence as a function of the model's potential} in a Press Schechter scheme to obtain the
differences in halo mass functions. The comparison is also done with
the more usual scenario $\Lambda$CDM. We conclude \textbf{that the
key role lies within the evolution of the equation of state and that
the method displays promisses for discrimination
using cluster mass determination by upcoming surveys}.
\end{abstract}

\noindent{\it Keywords\/:  semi-analytic modeling, dark matter, galaxy
clusters, dark energy theory
\\}
\submitto{J. Cosmology Astropart. Phys.}
\maketitle

\section{Introduction}

Recent evidence is piling up in favour of a dark energy component
in the dynamics of the universe forming up to 70\% of the energy density:
from SNIa magnitudes \cite{Riess98,Perlmut98}; and more recently
\textbf{\cite{WangMukh04,NesserisPeriv04,Riess04,DalyDjorgovski04,BiesiadaEtal04}};
from the CMB (Cosmic Microwave Background) measured by WMAP (Wilkinson
Microwave Anisotropy Probe) coupled with complementary inputs \cite{BennettEtal03,PageEtal03}
(for reviews, see \cite{Padman03,Sahni04}). The nature of this dark
energy is still unknown but could lie within two major families: it
is either coming from a static cosmological constant or developing
from a dynamical scalar field just emerging from subdominance called
quintessence \cite{RP}. The interest of quintessence lies in the
possibility it holds to solve the four major problems raised by the
cosmological constant model: the problem of a need for fine-tuning
the cosmological constant to unnatural values of vacuum energy, the
problem of coincidence between its present energy density and the
order of magnitude of the critical density, the problem of the equation
of state of the cosmological fluid, which can only take one value
for a cosmological constant, and the problem of building a model that
would naturally come up with solutions to these problems while coming
from fundamental physics. 

The drive that lead to quintessence models was based on the attempt
to address those four problems of the observed behaviour of dark energy.
The first problem requires a theory which yields the present value
for the dark energy \textbf{without the need to tune initial conditions
at unnaturally small values compared with} the natural energy scales
of the \textbf{early universe}. This addresses the huge discrepancy
(typically $M_{Pl}^{4}/\rho_{DE}\sim10^{120}$\textbf{; $M_{Pl}$:
Planck mass; $\rho_{DE}$: Dark Energy density}) 
commonly found between classical quantum vacuum energy calculations
and its cosmological measurements. The coincidence problem deals with
the current transition status of the universe between dominations
by matter and dark energy. It requires a theory that yields such a
tuned peculiar equilibrium without strong constraints on its initial
conditions. 
it deals with the value of the dark energy component $P/\rho$ ratio,
which is constrained to lie within $\left[\begin{array}{lll}
-1 & ; & -0.6\end{array}\right[$ \cite{Wangetal00} or even within $\left[\begin{array}{lll}
-1 & ; & -0.8\end{array}\right[$ \cite{Efst00,HannestadMortsell04} (for dark energy equation of state
supposed to respect the weak energy condition, thus $P/\rho>-1$).
Finally the model building problem express the need to have a theoretically
motivated dark energy potential.

The fine tuning problem has been the main drive for quintessence models,
hence it is solved by almost all proposed quintessence potentials,
involving high energy physics natural energy scales. Various models
have been proposed but few are deeply grounded in high energy physics.
The main examples of motivated potentials are \textbf{the original
Ratra-Peebles' potentials \cite{RP} that have been linked with the
context of global SUSY \cite{Binetruy99} and the so-called SUGRA
potentials derived from supergravity arguments \cite{BraxMartin99,BraxMartin00}}. The tracking
property is attracting much interest because it solves the coincidence
problem naturally. It defines classes of potentials which Klein-Gordon
equation admits a time varying solution that acts as an attractor
for wide ranges of field initial conditions. In \cite{Steinhardtetal99},
general conditions to obtain a tracking potential are given.

In front of the wealth of scalar fields available from physics beyond
the standard model, the need for discrimination between different
quintessential models proposed in the literature is a strong drive
to confront their predictions with observable features. Since the
advent of the COBE satellite results and the pursuit of refined Cosmic
Microwave Background Radiation (CMBR) measurements, the attention
of cosmologists dealing with quintessence had been focused on CMBR
anisotropy measurement as primary probes for the various quintessence
models (e.g. \cite{Brax00}). That has foreshadowed the possibility
offered by quintessential cosmic dynamics to alter in turn the formation
of large scale structures, that are more readily available to observations.
Some of the earlier attempts in this direction that have been started
out have restricted themselves to linear or perturbation theory of
structure formation \cite{BenabedBern01}, to pseudo-quintessence
models (approximation of a constant equation of state parameter different
than the $\Lambda$ term; \cite{LokasHoffman01,LokasEtal04,KuhlenEtal05})
or have not pushed the envelope beyond the study of the spherical
collapse model \textbf{\cite{MaininiEtal03b,MotaVdB00,NunesMota04}}
or the impact on halo concentrations \cite{DolagEtal04,KuhlenEtal05}. 

The aim of this paper is to compute the non-linear mass function of
collapsed structures in the presence of a quintessence field. One
approach is to use direct numerical simulation \cite{KlypinEtal03,LinderJenkins03,DolagEtal04,MaccioEtal04,KuhlenEtal05,SoleviEtal05},
but this is only practical for a few models with well defined parameters
and gives little physical insight into the influence of the quintessence
potential. Our approach takes into account the fully dynamical nature
of the field using the methodology developed by Press \& Schechter
\textbf{\cite[hereafter PS]{PS}}. This method uses a spherically symmetric
dynamical model to relate the collapse of massive structures to a
density threshold in the linearly extrapolated density field. In this
way, it is possible to apply Gaussian statistics to the initial density
field in order to count the numbers of collapsed structures above
a given mass threshold at a particular epoch. Because the method is
nearly analytic, it can be applied to rapidly expore the parameter
space of possible quintessence models and inverted to constrain quintessence
models using observational data. Some approaches using semi-analytical
methods have been published in the course of the present work with
a restricted range of quintessence potentials \cite{MaininiEtal03a,MaininiEtal03b,SoleviEtal04,NunesMota04},
some restricting to very indirect observables \cite{MaininiEtal03b,NunesMota04}.
None have yet looked at the same time at observables like the mass
function with many different potentials and an approach aimed at understanding
the dominant physics involved, using the full potential of semi-analytical
methods.

The paper is laid out as follows. In section~(\ref{sec:Homogeneous-quintessence-evolution}),
homogeneous universe models with quintessence are explored and familiarity
with the physics of quintessence is developed. Some results on the
behaviour of the universe's radial scale in various such models are
presented, showing the possibility offered by large scale structures
to probe the quintessence potentials. In section~(\ref{sec:Spherical-collapse-in}),
the collapse of a top hat spherical model for a single primordial
inhomogeneity is computed. This provides a simple way of linking perturbations
in the initial density field with collapsed structures. In section~(\ref{sec:The-mass-function}),
we combined this model with information from the statistics of the
initial inhomogeneities. Using a PS type scheme we obtain predictions
for the mass function in the presence of various quintessential models,
and compare the evolution of different models. Finally, in section~\ref{sec:Conclusion}
we summarise the results, and explore the constraints that future astronomical
surveys will be able to set on the form of the quintessence potential.

\section{Homogeneous quintessence evolution\label{sec:Homogeneous-quintessence-evolution}}

In this section we will review the features in the homogeneous models
of dynamical dark energy and emphasize that which points toward an
effect on structure formation. \textbf{This roadmap will later be
useful for interpreting our results.} The first step is to establish
the behaviour of various quintessence models within FLRW-type solutions.

\subsection{The dynamical system}

In a homogeneous model, the system evolution is entirely defined by
the evolution of its scale factor and its quintessence field with
its time derivative. The Einstein's Field Equations governing the
Friedman Lemaitre Robertson Walker universe and the Klein-Gordon Equation
for the quintessence homogeneous, time dependent,  scalar field can
be written as evolutions for the scale factor $a$ and the scalar
field $Q$ .

The first Friedman equation of our model can be expressed as\begin{eqnarray}
\dot{a}^{2} & = & a^{-1}\left(\Omega_{m_{0}}+\Omega_{r_{0}}a^{-1}+\frac{\rho_{Q}}{\rho_{c_{0}}}a^{3}+\Omega_{\Lambda_{0}}a^{3}+\Omega_{k_{0}}a\right),\label{eq:homoHubble:1}\end{eqnarray}

the dots refer to cosmic time derivatives and $^{\prime}$ will refer
(e.g. in Eq.\ref{eq:homoHubble:2} \textbf{below}) to the total derivative
with respect to the field Q. \textbf{Conventionally, subscript 0 refers
to the present epoch. $\Omega_{X}=\rho_{X}/\rho_{c}$ represents the
density parameter of species X, $\rho_{c}$ is the critical total
density required by a flat universe model. We introduce here the density
factor $\Omega_{X}^{\dagger}=\rho_{X}/\rho_{c_{0}}$; note $\Omega_{X_{0}}^{\dagger}=\Omega_{X_{0}}$. }

Each term corresponds respectively to the contribution in matter (dark
and luminous: $\Omega_{m_{0}}$), radiation (neutrinos and photons:
$\Omega_{r_{0}}$), cosmological constant ($\Omega_{\Lambda_{0}}$),
either time dependent ($\Omega_{Q}^{\dagger}=\frac{\rho_{Q}}{\rho_{c_{0}}}$)
or present epoch ($\Omega_{Q_{0}}=\rho_{Q_{0}}/\rho_{c_{0}}$, as
follows) quintessence density \textbf{factor} and the corresponding FLRW
curvature $\Omega_{k_{0}}$, with the definition\begin{eqnarray*}
\Omega_{k_{0}} & = & 1-\left(\Omega_{m_{0}}+\Omega_{r_{0}}+\Omega_{Q_{0}}+\Omega_{\Lambda_{0}}\right).\end{eqnarray*}
In this work we have assumed throughout a globally flat geometry of
the universe ($\Omega_{k_{0}}=0$) and a fully dynamical dark energy
($\Omega_{\Lambda_{0}}=0$).

It should be noted that the very non-linear Klein-Gordon equation
requires forward integration to reach the tracking solution while
the boundary conditions in the density parameters are set for a backward
time integration.

\label{sec:PlanckUnitsSubSubSec}

The natural units for a primordial quantum field like quintessence
are Planck units \textbf{(Planck time: $t_{Pl}$)}. In those units,
expressing the characteristic time\textbf{,} which is the Hubble time\textbf{:}
(\cite{Bahcall99,Freedman00} $t_{H_{0}}^{-1}=H_{0}\simeq65km\cdot s^{-1}Mpc^{-1}\simeq2.106×10^{-18}s^{-1}=(8.65\times10^{60}t_{Pl})^{-1}$)\textbf{,}
shows a discrepancy between the two timescales of the problem of more
than 60 orders of magnitudes. Since the Planck time should govern
the evolution of the field, this could have revealed problematic if
we were not focusing on cosmic evolution of the quintessence field:
neglecting the effects very of rapid fluctuations, we can observe
the evolution of the field as following a regular evolution of its
energy density if we adopt the Hubble time as characterizing our system. 

We also assumed spatial homogeneity of the field. \cite{Maetal99,LokasHoffman01}
argue for homogeneous Q, given that they found the smallest scale
of Q fluctuations to be larger than the clusters scale. Nevertheless
it has to be mentioned that their results hold their validity from
their studies in the mildly non-linear regime. \cite{MotaVdB00,NunesMota04,MaccioEtal04}
argue that the highly non-linear regime involved might require some
quintessence clustering or coupling to the dark matter. We will concentrate
here on effects in non-clustering minimally coupled models \textbf{as
a threshold model to ascertain the impact of quintessence on non-linear
clustering}.

In Eq.(\ref{eq:homoHubble:1}), where we also assume no curvature
nor cosmological constant contribution, the effect of quintessence
is focused in the density \textbf{factor}, involving the energy density
of the field $Q$ defined with its potential: in our units it reads
$\Omega_{Q}^{\dagger}=\frac{8\pi}{3}\left(\frac{\dot{Q}^{2}}{2}+\frac{V(Q)}{\mathcal{H}_{0}^{2}}\right)$
\textbf{following notations defined above and with $\dot{Q}$, the
quintessence field (Hubble) time derivative, $V(Q)$ its potential
energy and $\mathcal{H}_{0}$ the Hubble constant in Planck mass units}.
Its dynamics is governed by the Klein-Gordon equation in the case
of homogeneity:\begin{eqnarray}
\ddot{Q} & = & -3\frac{\dot{a}}{a}\dot{Q}-\mathcal{H}_{0}^{-2}V^{\prime}(Q).\label{eq:homoHubble:2}\end{eqnarray}
Though the pressure of the field is not involved in its homogeneous
evolution, it is crucial to the effect of quintessence on non-linear
collapse: for a scalar field recall that (with cosmic time)  $P_{Q}=\left(\frac{1}{2}{\dot{Q}}^{2}-V(Q)\right),$
and together with the density, they define the acceleration state
of the universe. One can characterise it using the equation of state\begin{equation}
\omega_{Q}=P_{Q}/\rho_{Q}\label{eq:EqnofState}\end{equation}
(see figure \ref{fig:RP6}'s upper panels). In terms
of the energy momentum tensor of the field, recall that for scalar
fields with these definitions of density and pressure, we have $\rho_{Q}\equiv T_{00};T_{j}^{i}\equiv P_{Q}\delta_{j}^{i}.$

Now, we will discuss a variety of potentials proposed for quintessence
models and their \textbf{previously known} homogeneous properties\textbf{,
emphasizing their impact on matter domination}.


\subsection{Explorations with several tracking potentials\label{sec:trackingPropSubSec}}

\begin{table}

\caption{Our choice of various potentials satisfying the \protect\cite{Steinhardtetal99}
test. They are discussed in section \ref{sec:PotDefsSubSubSec}. They
are all either of inverse power\textbf{, gaussian} or exponential
types.\textbf{$\alpha_{Q}$ and $\lambda$ are slope parameters, $\Lambda_{Q}$
characterises the potential's energy scale and the values of the field
$Q$ are in (dimensionless) units of Planck mass. $\kappa=8\pi G=8\pi/m_{Pl}^{2}$
is the gravitational coupling ($\kappa=8\pi$ in units of $m_{Pl}$).}\label{tab:TrackerTest}}

\begin{center}\begin{tabular}{p{4.5cm}l}
\br
Name &
Potential V \tabularnewline
\mr
R.P. \cite{RP}&
 $\frac{\Lambda_{Q}^{4+\alpha_{Q}}}{Q^{\alpha_{Q}}}$\tabularnewline
SUGRA \cite{BraxMartin00} &
 $\frac{\Lambda_{Q}^{4+\alpha_{Q}}}{Q^{\alpha_{Q}}}e^{\kappa\frac{Q^{2}}{2}}$\tabularnewline
\cite{FerreiraJoyce98} &
 $\Lambda_{Q}^{4}e^{-\lambda Q}$\tabularnewline
\cite{Steinhardtetal99}&
 $\Lambda_{Q}^{4}e^{\frac{1}{Q}}$\tabularnewline
\br
\end{tabular}\end{center}
\end{table}
In this section we restrict our choice upon a set of potentials and
recall their previously studied equation of state (Eq.\ref{eq:EqnofState})
and density parameter ($\Omega_{Q}$) homogeneous evolution, emphasizing
behaviours that can affect the formation of large scale structures.

We have narrowed our study on such potentials from the literature
that we found to agree (at least marginally) with the \cite{Steinhardtetal99}
general conditions to obtain a tracking potential. \label{sec:PotDefsSubSubSec}We
therefore selected several forms of potential (the list is not exhaustive)
\cite{RP,BraxMartin99,BraxMartin00,FerreiraJoyce98,Steinhardtetal99}
for the rest of this study. The Ratra-Peebles potential \textbf{\cite[hereafter RP or Ratra-Peebles]{RP}},
first discussed potential in the literature has retained its interest
with its more recent discussion within the context of global SUSY
\cite{Binetruy99}. The \textbf{\cite[hereafter SUGRA]{BraxMartin00}}
potential has been motivated in the framework of supergravity low
energy approximation. The simple exponential \textbf{\cite[hereafter FJ or Ferreira \& Joyce]{FerreiraJoyce98}}
potential displays a generic form for moduli fields from extradimensional
theories flat directions while the \textbf{\cite[hereafter Steinhardt \it et al.]{Steinhardtetal99}}
potential has been proposed as an infinite sum of Ratra-Peebles-type
potentials. This set of potential has been chosen as a well motivated
starting point. It also spans the main types of potentials, power
laws, \textbf{gaussian} and exponentials and display very different
behaviours.

Using those potentials, spelled out in table \ref{tab:TrackerTest},
we \textbf{can explore} the impact of different models on the timescale of
the quintessence dominated epoch and the apparent strength of matter
domination between radiation and quintessence eras. 

\label{sub:Comparing-potentials}We also can compare the observable
effects of various potentials at the homogeneous level through the
equation of state evolution. The equation of state evolution should
be attained by SNIa measurements which constrains directly the integrated
luminosity distance evolution, although it relies on SNIa to be good
standard candles without systematic errors. In this paper, we are
even more interested in the density parameters: since, on the rough,
dark energy with its negative pressure has a freezing effect on clustering
of dark matter, the differences between potentials in the equivalence
epoch for matter-quintessence and in the strength of the matter dominance
phase are signs respectively of differing inhibiting times for structure
formation and of overall matter clustering activity.

We thus emphasize those features in the otherwise known homogeneous
evolution with \textbf{Ratra-Peebles} potentials for the slope $\alpha_{Q}=6$
and $11,$with the \textbf{SUGRA} potentials for the slope
$\alpha_{Q}=6$ and $11,$ \textbf{Ferreira \& Joyce}
potential for \textbf{$\lambda=10$
}%
\footnote{\textbf{This choice is historical and does not allow the marginal
tracking behaviour of FJ to be reached. It allows nevertheless to
explore a very different behaviour of the equation of state that yields
crucial insights (see section \ref{sub:Integrated-mass-functions}).}%
}and the \cite{Steinhardtetal99} potential
(the choices for parameter values follow the authors). Comparison
of the density parameters evolutions allows to conclude on the fact
that we expect a stronger inhibition of structure formation with the
Ratra-Peebles potential than with the \cite{Steinhardtetal99}, than
with the SUGRA, than with the Ferreira \& Joyce potentials. Within
models (i.e. for Ratra-Peebles and SUGRA), \textbf{variations of their respective evolution points} towards the degree of freedom inherent
to each potential and thus towards the falsifiability of structure
formation tests on each model. Another remark drawn from the density
parameters concerns the scales of the structures affected: in the
hierarchical CDM scenario, clusters and superclusters scales being
formed last, we expect them to be most affected by the changes in
inhibition epochs from the various potentials because those epochs
occur during the most recent periods and inhibition is expected to
\textbf{act most on} them. Eventually it should be stressed that the observational
constraints being applied nowadays, differences between models are
expected to increase as we look back in time. It should be noted that
we do not expect subdominant quintessence to alter the matter radiation
equivalence, and in relation, the recombination, thus the power spectrum
to remain essentially unchanged for the purpose of structure formation.

\begin{figure*}
\begin{center}\includegraphics[%
  clip,
  width=1.0\textwidth,
  keepaspectratio]{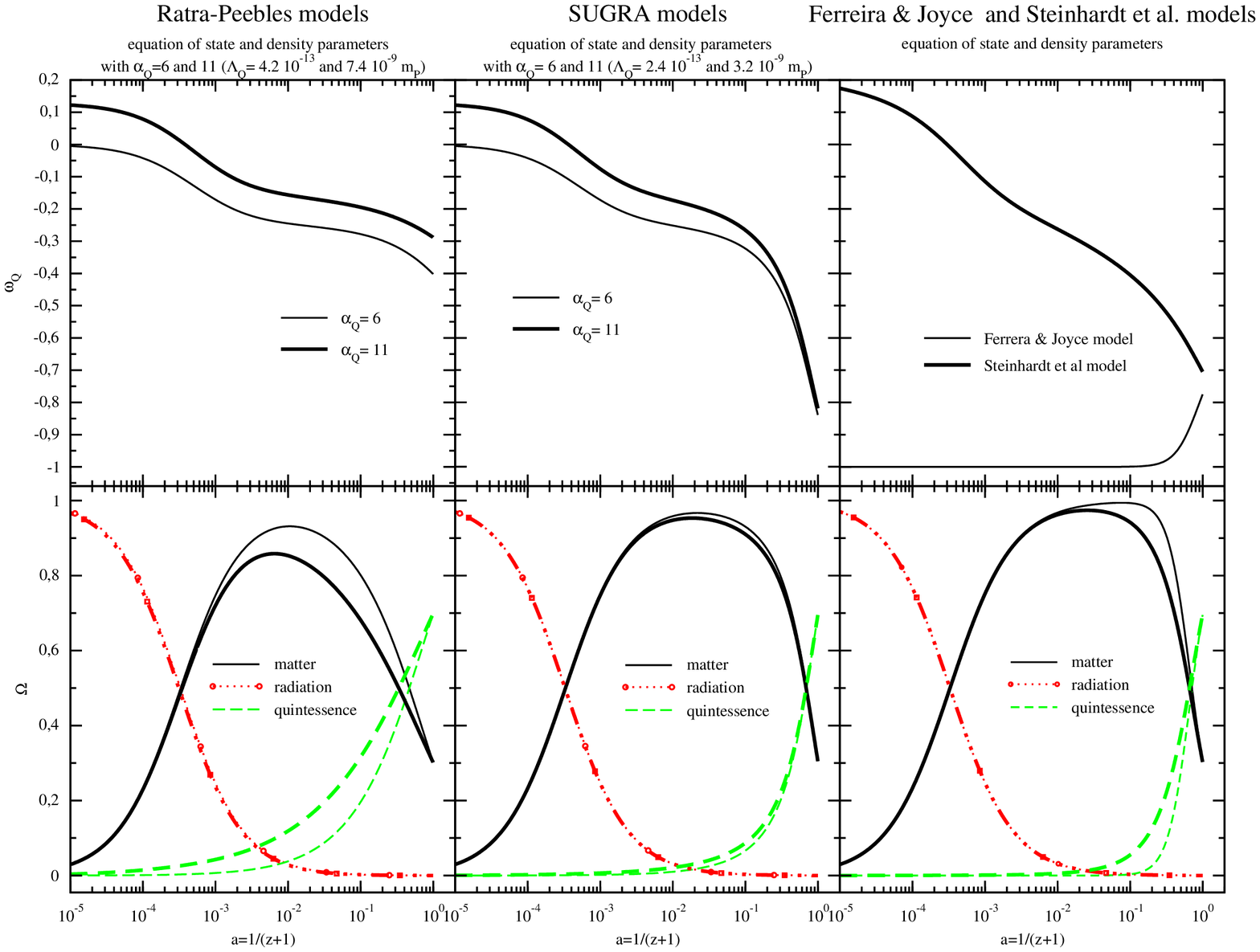}\end{center}

\caption{\label{fig:RP6}\label{fig:RP11}Calculations using the Ratra-Peebles
\cite{RP} and \label{fig:SUGRA6}SUGRA \cite{BraxMartin00} models
for the shape parameter values $\alpha_{Q}=6$ and $11$, the Ferreira
\& Joyce \cite{FerreiraJoyce98} and \cite{Steinhardtetal99} models\label{fig:FJ}.
The equations of state $\omega_{Q}$ of all models are definitely
not constant over structure formation era. The \cite{FerreiraJoyce98}
model, although constant for a while, also varies strongly at recent
epochs. The density parameters show (i) matter-radiation equivalence
is unaffected by subdominant quintessence, (ii) changes between models
are backwards since they are jointed by present density conditions,
(iii) models differ in matter quintessence equivalence (which late
epoch poses the so called coincidence problem) and (iv) models differ
in the strength of matter dominance. For both Ratra-Peebles models,
these discrepancy illustrate the variability of this potential and
relativize its testability. For both SUGRA models, they illustrate
the narrow variability of this potential and emphasize its testability,
although distinguishing between the shape parameters may prove difficult.
The equations of state shows that the models are well accommodating
the measurements \cite{Wangetal00,Efst00}, except for the Ratra-Peebles
models which is their know pitfall. }
\end{figure*}

\textbf{The lower panels of figure \ref{fig:RP6} show} some changes for the Ratra-Peebles model that are much less
pronounced in the case of the SUGRA model,
and also illustrate\textbf{,} together with \textbf{the
FJ/Steinhardt} \textbf{\emph{et al.}} \textbf{panel,} the variations
in equivalence epochs \textbf{and strength of matter domination}. The upper panels all illustrate the fact that
the equation of state cannot be considered constant during the epoch
of structure formation (contrary to \cite{LokasHoffman01}).

\label{sub:Numerical-implementation-Homog}The solving of Eqs.(\ref{eq:homoHubble:1},
\ref{eq:homoHubble:2}) was effected with a second order Runge-Kutta
integration method. As previously mentioned, giving initial conditions
for each potential shown in table \ref{tab:TrackerTest} is not constrained
with input observations. The search for tracking solutions excludes
reverse integration methods: the selection of the tracker solution
which leads to observed quintessence density was obtained with a simple
iteration on the characteristic energy scale $\Lambda_{Q}$ with respect to the target
final quintessential density.

The cosmic time numerical increment is chosen so as to keep a constant
logarithmic scale factor increment: $dt=\alpha\frac{a}{\dot{a}},$
where $\alpha$ is fixed (we usually take $\alpha=10^{-2}$, a value
of $10^{-3}$changing the outcome by less than 1\%).

Given the \textbf{roadmap} that we \textbf{now} have on variations
\textbf{between models} in the influence of quintessence on large
scale structure formation\textbf{. We are going}
 to effectively test them in the mass functions of
cosmological haloes within quintessential context as computed by the
PS scheme \cite{PS}. First, in order to get the mass function, we
need to study the spherical collapse model. This is done in the next
section.


\section{Spherical collapse in the presence of quintessence\label{sec:Spherical-collapse-in}}

The influence of quintessence models on structure formation comes
from the repulsive gravitational effect of its negative pressure on
the bulk of spacetime during its phase of less than -1/3 equation
of state. Collapsing structures are then slowed down in their build
up by this relative aggravated expansion.  An elementary model of
this phenomenon can be used to get simple quantitative results that
can then be inputted in a PS-type scheme that will be discussed in
section \ref{sec:The-mass-function}: the spherical collapse model
(pioneered in \cite{Larson69,Penston69,GG72,FG84,Bertsching85} and
summarised in \cite{Peebles}) in the presence of quintessence holds
the key to this exploration from the beginning of the collapse phase
after recombination down to shell crossing. After shell crossing,
mass conservation does not allow to follow the system with just its
outer shell but a prescription can be used for the virialization of
the model. We will now describe the dynamics of the cosmological spherical
collapse.


\subsection{The dynamical system}

The cosmological spherical collapse is embedded into an FLRW-type
universe for which all the component of energy density are supplemented
with a spherical overdense region. Birkhoff's theorem holds the key
to the spherical non linear collapse model: any spherical region embedded
in a spherically symmetric universe behaves shell by shell as a patch
of FLRW universe with each characteristics modified to match the corresponding
average inner ones. In this case, any given shell behaves according
to the average of its inner density. With a flat background this leads
to positive curvature inside the overdensity. However, because this
inner curvature may change with time, one has to be cautious in using
the Friedmann's equations \cite{MotaVdB00}.

For simplicity reasons and because it is a building block of the original
PS scheme, we will use the spherical top hat model (sphere of constant
overdensity). In this case, the averages are equal to the \textbf{local
values} and
every shell reach the center of the overdensity at the same time,
when shell crossing occurs. Problematic for the virialization, this
feature allows one to follow only the evolution of the outermost shell
representing the whole system. Virialization is then assumed to take
place soon after shell crossing and a halo of mass given by the extent
of the initial overdensity is formed.

Thus the radius \textbf{$r_{od}=r$} of the overdensity can be written
\textbf{as a function of its initial value $r_{i_{od}}=r_{i}$} as $r_{od}(t)=r_{i_{od}}a_{k>0}(t)=r_{i}s(t),$
where we note the scale factor of the positive curvature patch \textbf{$a_{k}>0$} as
a fiducial or rescaled radius of the overdensity region $s=r/r_{i}.$
Thus the radius of the overdensity follows Friedman's equation for
the modified overdensity patch of universe. For this shifted FLRW
model, the Einstein's Equations yield the evolution \textbf{(recall $\kappa=8\pi G$, the
gravitational coupling)}:\begin{eqnarray}
\dot{s}^{2} & = & \frac{\kappa s^{2}}{3}\sum\rho-k(t),\label{eq:FEcosmic}\\
\ddot{s} & = & -s\left[\frac{\kappa}{6}\left(\sum\left(3P+\rho\right)\right)\right]\label{eq:FAcosmic}\end{eqnarray}
where the sums on energy density and pressure concern each cosmic
species inside the patch radius. Again note that the curvature \textbf{$k$} inside
the overdensity is not constant in general, so that we prefer to use
Eq.(\ref{eq:FAcosmic}). Since we assumed homogeneous behaviour for
the scalar field (i.e. no clustering), that means that there is no
conservation inside the spherical patch for the quintessence field
\cite{MotaVdB00} and \textbf{its} pressure is that \textbf{for} the background universe.

The acceleration follows the shifted second Friedman equation (Eq.
\ref{eq:FAcosmic}), which contains as well the pressure terms. The
density term for the matter follows the same pattern as for Eq. (\ref{eq:FEcosmic}):
\[
-\frac{\kappa r}{6}\rho=-\frac{4\pi Gr}{3}\rho=-\frac{GM}{r^{2}},\]
 with the total mass M conserved and contained initially inside the
spherical patch. The pressure terms depend on the respective state
equations. Hence the acceleration equation reads, with our time and
radius units, \begin{eqnarray}
\ddot{s} & = & \left[\lambda_{0}-\left(\Omega_{r_{0}}a^{-4}+\frac{4\pi G}{3\mathcal{H}_{0}^{2}}(3P_{Q}+\rho_{Q})\right)\right]s-\frac{\Omega_{m_{0}}a_{i}^{-3}\left(1+\Delta_{i}\right)}{2s^{2}}.\label{eq:SphCollPlanckAcc}\end{eqnarray}

Initial conditions of the homogeneous evolution are taken after inflation
($a=10^{-30}$), the field is taken in a reasonable range allowing
for the tracking solution to establish. For the collapse evolution,
initial time is chosen in the relevant overdensity linear regime (we
usually take the arbitrary cut \textbf{$a_{i}=10^{-5}$}), the overdensity is set
in a Hubble flow, that is following the general expansion of the universe
at the initial onset of the overdensity, (the definition of $s$ sets
its initial condition to be \textbf{$s_{i}=1$} at initial time) so\textbf{\begin{eqnarray}
\dot{s}_{i} & = & \frac{\dot{a}_{i}}{a_{i}}.\label{eq:initODflow}\end{eqnarray}
}


%
\begin{figure*}
\begin{center}\includegraphics[%
  clip,
  width=0.50\columnwidth,
  height=1.0\textwidth,
  keepaspectratio]{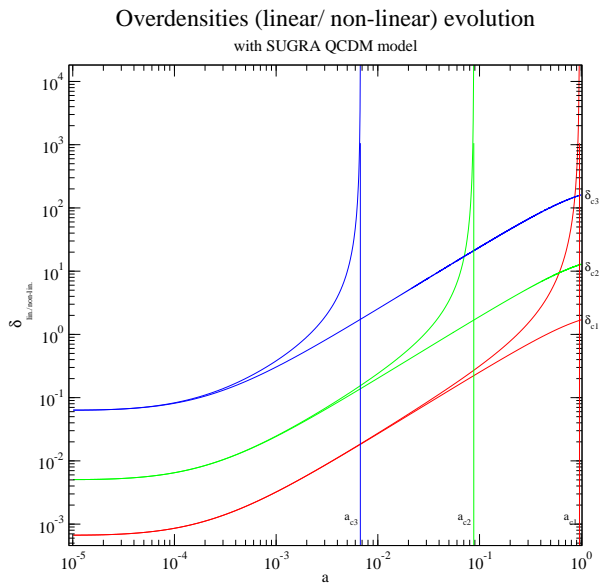}\end{center}

\caption{\label{cap:oneCollapse} \label{cap:rangeCollapse}Linear and non-linear
overdensity evolutions for a range of initial overdensities illustrating
the computation of the $\delta_{c_{0}}(a_{c})$ function for the Ratra-Peebles
model and several values of $\Delta_{i}$. The non-linear collapse
ends with diverging overdensity at the corresponding collapse scale
$a_{c}$ while from the linear evolution, initial condition $\Delta_{i}$
is made to correspond to its present day linearly extrapolated value
\textbf{$\delta_{c_{0}}$.}}
\end{figure*}

\textbf{In the model of the spherical collapse, the non-linear density
at the border of the overdense region can be monitored using the Lagrangian
mass conservation. For the calculation of the overdensity, one just
follows the canonical definition $\delta=\frac{\rho_{m}-\rho_{b}}{\rho_{b}}$,
then\begin{equation}
\delta=\left(1+\Delta_{i}\right)\left(\frac{a}{a_{i}s}\right)^{3}-1.\label{eq:NLoverdensity}\end{equation}
}

\textbf{The} corresponding \textbf{initial} overdensity value thus
proceeds from Eq.(\ref{eq:NLoverdensity}) at $a=a_{i}$ and $s=1$
\textbf{and using Eq.(\ref{eq:initODflow})}:\begin{eqnarray}
\delta_{i} & = & \Delta_{i},\\
\dot{\delta_{i}} & = & 3(\delta_{i}+1)\left(\frac{\dot{a}}{a}-\frac{\dot{s}}{s}\right)_{i}=0.\end{eqnarray}

Here we have used a fiducial initial time from which the collapse
was evolved. This choice is of course arbitrary and in the PS scheme,
this arbitrariness is resolved by the extrapolating to present time
(redshift $z=0$) the initial overdensity $\delta_{i}=\Delta_{i}$
using the linear evolution theory\begin{equation}
\ddot{\delta_{L}}+2\frac{\dot{a}}{a}\dot{\delta_{L}}=\frac{3}{2}H_{0}^{2}\Omega_{m_{0}}a_{i}^{-3}\delta_{L}.\label{eq:LinDenEvol}\end{equation}
This theory is only valid for overdensities in the linear regime ($\delta\ll1$)
but the non-linear collapse leads to infinite density at a finite
time given by $a=a_{c}$ while the linear theory allows the density
to remain finite up to $z=0$.\label{sub:Construction-of-delta_c(a_c)}

The immediate interest of solving this model lies in the possible
comparison between different collapse times. The relevant quantity
extracted was the function $\delta_{c_{0}}=f(a_{c}),$ the linearly
extrapolated overdensity as a function of collapse epoch expressed
in terms of the scale factor. Given a potential, we construct its
characteristic extrapolated density contrast function of collapse
scale: each initial overdensity yields both the collapse scale and
extrapolated linear scale using the non linear and linear collapse
respectively (see figure \ref{cap:oneCollapse}). The construction
of the function involved scanning a whole range of initial overdensities
as illustrated in figure \ref{cap:rangeCollapse}. The critical density
contrast function of collapse scale factor $a_{c}$ is constructed
by the linear extrapolation to the present from the arbitrarily chosen
initial epoch's starting overdensity. Therefore $\delta_{c_{0}}$
yields an object that collapses at an epoch given by the value of
scale factor $a_{c}$. Thus $\delta_{c_{0}}$ is found by evolving
our initial conditions with Eq.(\ref{eq:LinDenEvol})'s linear theory,
and the corresponding collapse scale is found by the non-linear collapse
of Eq.(\ref{eq:SphCollPlanckAcc}) -- given by the vertical asymptote
at $a_{c}$, in the homogeneous background provided by Eqs.(\ref{eq:homoHubble:1},
\ref{eq:homoHubble:2}).

For implementing this scheme, we first determine the correct QCDM
potential energy scale, as described in section \ref{sub:Numerical-implementation-Homog},
then we bring the model to the field's tracking regime and produce
initial parameters for the spherical collapse. Eventually the coupled
evolution of the background FLRW, the quintessence field, the non-linear
spherical collapse and the linear overdensity evolution obtains the
values of $\delta_{c_{0}}(a_{c}).$

A fourth order Runge-Kutta integration method was implemented over
the whole Eqs.(\ref{eq:homoHubble:1}, \ref{eq:homoHubble:2}, \ref{eq:SphCollPlanckAcc},
\ref{eq:LinDenEvol}) system. We still used the logarithmic increment
$dt=\alpha\frac{s}{\dot{s}},$ where we usually take $\alpha=10^{-2}$,
as described in section \ref{sub:Numerical-implementation-Homog},
but we had to limit the lower increment value of $dt$ to avoid inflation
of numerical expenses when integration approaches the overdensity
turnaround point.

We are now ready to use the Top Hat Spherical collapse model to decide
when objects are considered to have collapsed, that is when their
non-linear overdensity diverges. The following section discuss those
first results.

\subsection{Critical densities\label{sub:The-results}}

\begin{figure*}
\begin{center}\includegraphics[%
  clip,
  width=0.50\columnwidth,
  height=1.0\columnwidth,
  keepaspectratio]{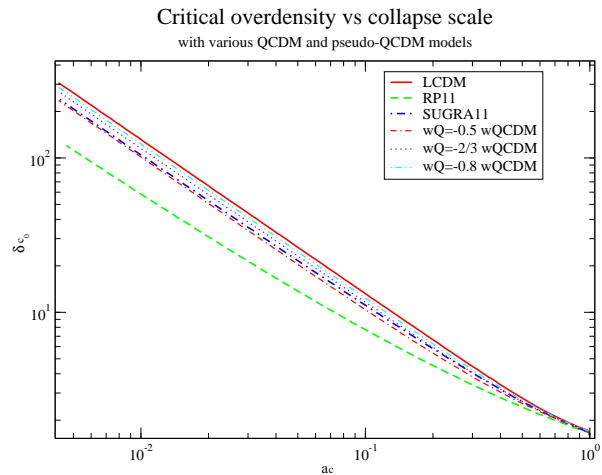}\end{center}

\caption{\label{cap:deltac0_ac2} Comparison between quintessence and pseudo-quintessence
models through their influence on non-linear collapse scale\textbf{,}
function \textbf{for the} linearly extrapolated overdensity. The notations
are following table \ref{tab:TrackerTest} with the number referring
to the index value used in the power of the potential \textbf{(e.g.
RP11: Ratra-Peebles with $\alpha_{Q}=11$)}. Pseudoquintessence models
are referred to by the value of their ad hoc equation of state. The
spread of curves shows possible distinctions between models. The crossing
of pseudo-quintessence models by SUGRA11 shows none can mimic its
evolution. LCDM, or $\Lambda$CDM, the boundary cosmological constant
model, is shown for reference. }
\end{figure*}
\begin{figure*}
\begin{center}\includegraphics[%
  clip,
  width=0.50\columnwidth,
  height=1.0\columnwidth,
  keepaspectratio]{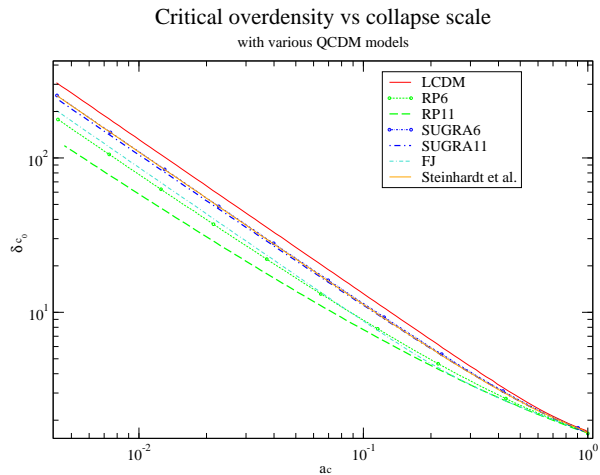}\end{center}

\caption{\label{cap:deltac0_ac1} Comparison between quintessence models through
their influence on non-linear collapse scale as a function of linearly
extrapolated overdensity. The notations are following table \ref{tab:TrackerTest}
with the number referring to the free index value used in the power
of the potential. The spread of curves shows possible distinctions
between models \textbf{(e.g. SUGRA6: SUGRA with $\alpha_{Q}=6$)}. LCDM, or $\Lambda$CDM, the boundary cosmological
constant model, is shown for reference.}
\end{figure*}

 Once we have computed a series of $\delta_{c_{0}}(a_{c}),$ for a
range of initial conditions\textbf{, set in the linear regime,} at
our arbitrary starting epoch, we can display for each model their
extrapolated critical overdensity as a function of \textbf{collapse}
scale factor and compare them among models, which is done in figures
\ref{cap:deltac0_ac2} (comparing pseudo-quintessence models) and
\ref{cap:deltac0_ac1} (for dynamical quintessence). 

Since we are dealing with density contrasts over the background cosmological
matter density, the main evolution effects are expected to come from
the lower \textbf{homogeneous} matter density yielded by the quintessence
models during their matter dominated era. Also the observation setting
the models corresponding to z=0 \textbf{($a=\frac{1}{1+z}=1$; present
epoch)}, we expect the differentiation between models to increase
backwards in time.

A comparison between our various potentials, the $\Lambda\mathrm{CDM}$
model%
\footnote{It should be noted that for a standard CDM \textbf{($\Omega_{m}=1$)} model,
the spherical collapse yields a linear evolution $\delta=\delta_{i}a/a_{i}$\cite{Peebles}
so, extrapolating the collapse value nowadays, $\delta_{c_{0}}=\delta_{c}1/a_{c}$
which gives the straight line asymptote to each models in these log-log
coordinates and the well known $\delta_{c}=1.686$.%
} and three pseudo-quintessential models (i.e. with constant equation
of state) have been performed. The overdensity as a function of collapse
shows clear distinctions between models are possible. The results
obtained are plotted on figures \ref{cap:deltac0_ac2}, for the pseudo
quintessences and two common models (Ratra-Peebles and SUGRA), and
\ref{cap:deltac0_ac1}, for all our studied models. 

The pseudo-quintessence models are shown for reference and comparison
to other works even though our homogeneous study has pointed that
such an assumption is highly unlikely to hold compared to real quintessence
models (e.g. \textbf{figure \ref{fig:RP6}'s upper} panels). However figure \ref{cap:deltac0_ac2} reveals that
even though the region scanned by the pseudo-quintessence models is
covering most of the values taken by our sample of models (except
for Ratra-Peebles), there is still a strong distinction if measurements
are confronted at different epoch since no pseudo-quintessence model
is matching quintessence evolution; for instance, the SUGRA11 (notation
defined in the figure caption) model cannot be fitted with only one
wQCDM model. 

The main point is that there are distinctions in this type of representations
between the various models when examined at earlier epochs and between
successive era. Whilst evaluation of variability for the SUGRA model
yields very \textbf{little separation}, that of the Ratra-Peebles are so large
that it allows for an observational selection of the free power index.
Whereas the FJ model crosses the \textbf{curves} from various values of
the free power index in the Ratra-Peebles model, the \cite{Steinhardtetal99}
model is not very distinguishable from the SUGRA6. However we will
see that the sensitivity of the mass function to the overdensity allows
for some distinction. 

To summarise, it is the departure, in quintessential spherical collapse,
from the constant value of the collapse critical overdensity of the
standard CDM collapse that will affect the mass function. On figures
\ref{cap:deltac0_ac2} and \ref{cap:deltac0_ac1}, this is translated
into the departure from the simple linear log-log relation of the
critical overdensity with the scale factor. Thus the linear density
contrast indexed by the non-linear scale of collapse sets the redshift
evolution of the PS mass function observable as we will see in the
next section.%

\section{The mass function in presence of a quintessence field\label{sec:The-mass-function}}

In this section we will lay down our assumptions for constructing
the mass dispersion of our models which imprints initial conditions
in the PS scheme, we will then recall it before discussing the integrated
mass functions for the quintessential models studied.

\subsection{Mass dispersion}

Initial conditions of large scale structure formation are condensed
in the PS scheme into their mass dispersion. To do this for \textbf{gaussian}
fluctuations the basic information lies in the density field power
spectrum.

In general, the Fourier power spectrum P(k) for the density fluctuations
\textbf{of scale $\frac{1}{k}$} can be decomposed into a primordial part and a linear evolution part,
written as follows (\cite{ColeLucchin95}pp 267,282, \cite{Peebles}p169,
see discussion in section \ref{sub:Linear-power-spectrumIn-the-case}):

\begin{equation}
P(k,a)=Bk^{n}T_{Q}(k,a)^{2}=Bk^{n}T_{\Lambda}(k,a)^{2}T_{Q/\Lambda}(k,a)^{2}\label{power}\end{equation}

where B is a normalization factor and $T_{Q}$ the \textbf{quintessence}
transfer function written as $T_{Q}=T_{\Lambda}$. $T_{Q/\Lambda}$\textbf{,
$T_{\Lambda}$ being the $\Lambda$ dark energy transfer function.}
The transfer function is itself written as: T$(k,a)=$T$(k,a=a_{i}).D(a)/D(a_{i})$
where the linear growth D gives the time evolution. 

In the following section, we briefly decompose the construction of
the structure formation initial spectrum between a primordial part
and a linear evolution, we discuss our assumed spectrum and show the
mass dispersion and its normalisation to observations.

\subsubsection{primordial power spectrum}

In the context of inflation, for commonly used potentials, one gets
a \textbf{gaussian} primordial spectrum P(k,$a_{init.}$)=$\left|\delta(k,a_{init.})\right|^{2}$=A$\lambda_{infl.}k$
(where $a_{init.}$ is the \textbf{epoch} at which a perturbation
of size l=1/k is reentering inside the horizon): i.e the scale invariant
Harrison-Zeldovich spectrum, with $\lambda_{infl.}$ and A constants
depending the inflationary potential and expansion period.

\subsubsection{Linear power spectrum}

To describe the growth of the fluctuations and the evolution of the
initial power spectrum P(k, $a_{init.}$) one usually evolves linearly
fluctuations up to a given \textbf{epoch $a$} (typically after recombination
when $\frac{1}{a}-1=$z$<$10$^{3}$) as P(k,t)=P(k, $a_{init.}$).T$^{2}(k,a),$
where T is the above transfer function.

Previous studies have shown that P(k) depends only weakly on the equation
of state $w_{Q}(a)=P(a)/\rho(a)$ \cite{LokasHoffman01} since the
Q field is sub-dominant at epochs where z is large (e.g., at recombination,
z$\simeq$ 10$^{3}$
). In the case of \textbf{a universe} dominated by cold dark matter
(CDM) the initial T(k) function is known to be well fitted \cite{BBKS,Sugiyama95}.
The $\Lambda$CDM (case $w_{Q}$=-1) transfer function T$_{\Lambda}$
fit is very close to the CDM function with a slightly smaller cutoff
scale \cite{EfstEtal92} and identical power laws. Flat cosmologies
transfer functions fits from N-body simulations of pseudoQuintessence
models (\textbf{!real bf!models with constant equation of state} $w_{Q}$)
have been proposed in the form\label{sub:Linear-power-spectrumIn-the-case}
$T_{Q}=T_{\Lambda}$. $T_{Q/\Lambda}$ \cite{Maetal99}. They show
only modifications to $T_{\Lambda}$ of order unity in the observationally
relevant range of $w_{Q}$.

Following previous work, we therefore apply the same initial density
fluctuation power spectrum to all of the models. This ensures that
the differences that we see result from the dynamical evolution of
the density fluctuations rather than differences in the initial power
spectrum. In practice, CMBR experiments will establish the fluctuation
power spectrum independently of the mass functions that we consider
here and a successful model will have to successfully reproduce both
datasets. We thus consider that the T$_{\Lambda}$ function given
for a model with a cosmological constant can describe suitably the
fluctuations and do not allow the Q field to influence the initial
power spectrum of density fluctuations. We \textbf{therefore} adopt
the \textbf{\cite{BondEfstathiou84}} form for the power spectrum\textbf{,
following our use of the \cite{Jenkinsetal01}}%
\footnote{\textbf{note their code was built to encompass $\Lambda$CDM dark
energy only}%
} \textbf{mass dispertion and mass function code, modified with the
present work's quintessential non-linear collapse program.}

\subsubsection{mass dispersion}

The mass dispersion $\sigma^{2}(M,a)$ is computed using a top-hat
filter in real space with a radius of filtering R, corresponding to
$M=4\pi\rho_{b}(a)R^{3}/3,$ $\rho_{b}(a)$ being the density of mass
of the Universe at a given epoch \textbf{marked with the scale factor
$a$} (thus we can express $R\left(M\right)=\left(3M/4\pi\rho_{b}(a)\right)^{1/3}$.

One has, for the filtered mass dispersion, the relation:\begin{equation}
\sigma_{Q}^{2}(R,a)=D^{2}(a)\int d^{3}k\left|W(R(a),k)\right|^{2}P(k)/8\pi^{3}\end{equation}
with the filter Fourier transform W (which is here the top hat filter),
indices in $Q$ denotes the model dependence. Thus we get the relation:\begin{equation}
\sigma_{Q}^{2}(R,a)=\sigma_{Q}^{2}(f(a)M^{1/3},a)=D^{2}(a)\int dk\left|W(R,k)\right|^{2}P_{BondEfsthathiou}(k)/8\pi^{3}\label{var}\end{equation}
with $f(a)=(3/4\pi\rho_{b}(a))^{1/3}$ and D, the linear growing
mode for fluctuations.

\subsubsection{spectrum normalization}

To get the value of the constant in P(k,a) one usually normalizes
$\sigma$ to the value $\sigma_{8h^{-1}}$ which is observed today
for a typical radius R= 8h$^{-1}$Mpc (here we take h=0.65) and to
reproduce the present amplitude of the density contrast one gets that
$\sigma_{8h^{-1}}$ is of order 1 (see for example reference \cite{Eke96}
that yields the product $\sigma_{8h^{-1.}}$ $\Omega_{0}^{1/2}=0.6,$
or $\sigma_{8h^{-1}}=$1.1 for $\Omega_{0}=0.3$).

\subsection{The mass function evaluation}

To compute the mass function in the non-linear regime without using
fits to N-body simulations we restrict here to the popular PS prescription.

This semi-analytical tool turns out to be the simplest way to construct
a mass function from the assumed \textbf{gaussian} statistics of the initial
density. It neglects a large number of effects that tend to compensate
each other yielding often accurate fits to the numerical simulations.

It is yet an empirical recipe yielding the number of (collapsed) objects
with a mass greater than a given one \cite{PS}.

Hence, we obtain the fraction of collapsed mass linearly extrapolated
at a selected \textbf{present} time t$_{0}$ defined by $\delta_{c_{0}}(t_{c})$ as
in section (\ref{sub:Construction-of-delta_c(a_c)}\textbf{),} and
account for Eq.(\ref{var}) for the variance $\sigma$ itself linearly
extrapolated (and normalised with $\sigma_{8}$) at the same time
t$_{0}.$ The mass fraction thus writes:

\begin{equation}
F(m>M,a)=1-Erf(\delta_{c_{0}}(a)/2\sigma(M,a)),\end{equation}

and its derivative with respect to the mass leads to the density of
collapsed objects

\begin{eqnarray}
n_{PS}(m,a) & = & -\rho_{b}(a)\left(dF(m>M,a)/dM\right)/M\\
 & = & (2/\pi)^{1/2}\left(\rho_{b}(a)\delta_{c_{0}}(a)\right)\left(d\sigma(M,a)/dM\right)/\left(M\sigma^{2}(M,a)\right)\nonumber \\
 &  & .\exp\left(-\delta_{c_{0}}^{2}/(2\sigma^{2}(M,a))\right)\\
 & = & \left(-\rho_{b}/\sigma M\right)\left(d\sigma(M,a)/dM\right)F_{PS}(M)\end{eqnarray}
with the characteristic for the PS scheme condensed into the function

\begin{equation}
F_{PS}(M)=(2/\pi)^{1/2}\left(\delta_{c_{0}}(a)/\sigma(M,a)\right)\exp\left(-\delta_{c_{0}}^{2}(a)/(2\sigma^{2}(M,a))\right)\end{equation}
that can be replaced in later studies with more complex schemes.

We now have set in place the machinery for computing mass functions
for non-linear structures in the presence of non-clustering dynamical
quintessence. We will then present comparisons of integrated mass
functions that can be obtained for the potentials we selected in the
following section.

\subsection{Integrated mass functions\label{sub:Integrated-mass-functions}}

\begin{figure*}
\begin{center}\includegraphics[%
  clip,
  width=1.0\textwidth]{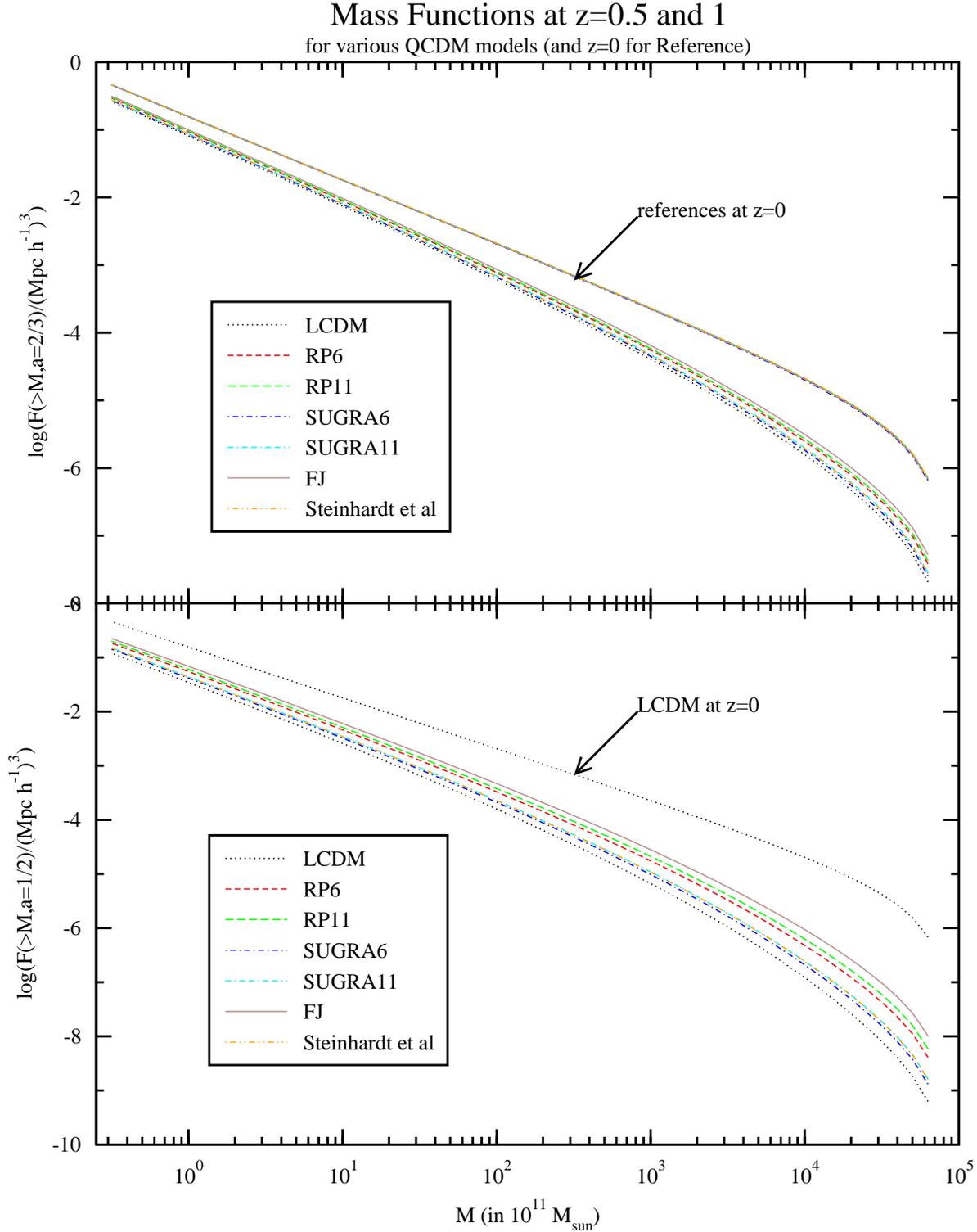}\end{center}

\caption{\label{cap:MassFunctionZ0}\label{cap:MassFunctionZ0.5}\label{cap:MassFunctionZ1}Computation
of the integrated mass function for various quintessence potentials
for past epochs, z=0.5 (a=2/3) and z=1 (a=1/2), compared with the
present epoch (z=0). The upper panels lack of dispersion among mass
functions for all models at z=0 illustrates our normalisation 
and the accuracy of our method. The spreads for past epochs, even
more so at the earliest epoch, allows for confrontation with cluster
mass measurements.}
\end{figure*}
One of the main interests of this study is to have shown the sensitivity
of $\delta_{c_{0}}$(a) as function of the chosen Q field potential
V(Q). This can be seen above (section \ref{sub:The-results}) with
figures \ref{cap:deltac0_ac2} \textbf{and} \ref{cap:deltac0_ac1}
when dealing with the collapse of an initial overdensity. Placing
these results within the PS scheme, we have found that some differences
among the various models can emerge from the mass functions. In figure
\ref{cap:MassFunctionZ0}, the upper panel shows as a reference the
mass functions for all the selected models at present time (z=0).
The models there are all very close to each others, which reflects
our z=0 normalisation and confirms the usual hypothesis taken of a
normalization of the mass functions to present days observations.
This is why figure \ref{cap:MassFunctionZ0}'s lower panel only \textbf{recalls}
the LCDM (\textbf{i.e.} $\Lambda\mathrm{CDM}$) mass function as a
reference. 

When we examine the mass function at larger z, as shown on figure
\ref{cap:MassFunctionZ0.5}'s upper panel for z=0.5 and lower panel
for z=1, discrimination can occur between the various models. All
models\textbf{' mass functions} are lying above the LCDM (\textbf{i.e.}
$\Lambda\mathrm{CDM}$) curve \textbf{for the same epoch}. The distinctions
remain nevertheless mild at z=0.5, as shown by the relative \textbf{proximity
between} the curves\textbf{; they}, however, become more pronounced
at z=1, with the following \textbf{observations}: the FJ model displays
the most structure suppression\textbf{, that is the least evolution
of mass function}, i.e. the least structure formation during recent
times. Then\textbf{, in the same decreasing hierarchy, one encounters
the} Ratra-Peebles models, in decreasing order of their power law,
the Steinhardt \emph{et al.} model, the SUGRA models, also in decreasing
order of their power law, before reaching the $\Lambda{\rm CDM}$
model (the Steinhardt \emph{et al.} and the SUGRA11 models\textbf{'
mass functions} are almost on top of each other). Surprisingly, those
results are not completely agreeing with the naive interpretation
given in section \ref{sub:Comparing-potentials}, that would have
yielded the same hierarchy of structure inhibition \textbf{and matter
dominance} except for the FJ model\textbf{, that would be expected
to have a} lower \textbf{mass function} than all other quintessence
models (see figure \ref{fig:FJ}'s lower panels). Confronted with
their lookback effects on the number of structures present at a certain
epoch, these are both agreeing with the mass function hierarchy except
for the ordering of the FJ model. We are thus compelled to consider
the only qualitative difference that springs to the eye between the
FJ model and the rest of them: the evolution of the equation of state
(figure \ref{fig:FJ}'s upper panels). The FJ model appears as the
only one \textbf{with $\omega_{Q}$} evolving towards higher values
at its latest stages, \textbf{during} structure formation. It seems
thus that the effect of its negative pressure in the acceleration
equation of the spherical collapse (Eq. \ref{eq:SphCollPlanckAcc})
is strong enough to counter the \textbf{strength of matter domination
earlier than expected in the} natural first approach. \textbf{This
can be seen from the ratio of quintessence- and matter-induced accelerations
of the overdensity, from Eq.(\ref{eq:SphCollPlanckAcc}), so we} get
$(3\left|\omega_{Q}\right|-1)\frac{\Omega_{Q}}{\Omega_{m}}>\frac{\left(1+\Delta_{i}\right)}{2(a_{i}.s/a)^{3}}$
despite having $\frac{\Omega_{Q}}{\Omega_{m}}$ small \textbf{at a
given stage of collapse and cosmic evolution}. These results cast
very interesting light into the intimate mechanisms of quintessence
that would escape to non semi-analytic approaches.

The evaluation of model's variability is expected after section \ref{sub:The-results}'s
more or less pronounced differences among each class of model and
their respective parameter variations (Ratra-Peebles and SUGRA). In
the light of integrated mass functions, \textbf{each model evaluated}
\textbf{displays little} variability so discrimination among power
law values may prove difficult with this method.

\section{Conclusions\label{sec:Conclusion}}

In this paper we have extended the evidence that several different
recent models of quintessence predict significant difference in the
evolution of the halo mass function, in general agreement with previously
more restricted works \cite{MaininiEtal03a,MaininiEtal03b,NunesMota04,SoleviEtal04},
and used our wider range of potentials to emphasize the impact of
the equation of state in the formation of dark haloes. Although the
best evidence seems to come from the largest mass structures, as expected
from the homogeneous exploration (section \ref{sec:PotDefsSubSubSec}),
the models studied do not behave simply according to \textbf{matter
and quintessence} dominance epochs. Nevertheless, the sensitivity
\textbf{of mass functions} to the value of $\delta_{c_{0}}$ \textbf{at}
the limits of accuracy cautions us against strong assertions. 

The models can be distinguished if observational measurements of the
cluster mass function at $z\sim1$ achieve a precision of better than
$\sim$10\% at $10^{14}\, h^{-1}M_{\odot}$. Because the models are
normalised to give the same mass function at $z=0$, a higher level
of precision, approaching 1\%, is required to distinguish between
the models at $z=0.5$. Several experiments are planned that will
exceed this level of accuracy and will have good control of the systematic
errors.

One of the most promising is the South Pole Telescope (SPT) survey
- a cosmic microwave background experiment aimed at detecting the
Sunyaev-Zeldovich effect (SZE) due to clusters of galaxies \cite{Ruhl04}.
The effect arises because inverse Compton scattering by the hot intra-cluster
plasma distorts the spectrum of the cosmic microwave background. Importantly,
the amplitude of the decrement is independent of the cluster redshift,
and the integrated signal decreases only with the clusters angular
diameter. The aim of the South Pole Telescope experiment is to map
a region of 4000 deg.$^{2}$, detecting essentially all clusters with
masses greater than $2\times10^{14}\, h^{-1}M_{\odot}$ and with redshift
less than 2 \cite{Carlstrom02}. The survey is expected to detect
around 30,000 clusters in the $\Lambda$CDM cosmology, with 30\% expected
to have redshifts greater than 0.8. The amplitude of the SZ decrement
is determined by the integrated pressure of the intra-cluster plasma,
and can thus be used to determine the cluster mass. Random errors
in the normalisation of the mass function will be $\sim1\%$ at both
$z\sim1$ and $z\sim0.5$. Systematic errors in the determination
of mass and the scatter between cluster mass and the observed SZ decrement
are a significant source of concern, but these can be controlled using
the ``self calibration'' techniques discussed in \cite{Hu03} and
\cite{Majumdar03}. An other technique based on the optical detection
of clusters may also yield promising results when combined with similar
self-calibration methods \cite{GladderYee05}.

These observational programmes will map the development of the
mass function from $z=1$ to $z=0$ at the level precision that is
required to distinguish between the quintessence models considered
in this paper. The techniques we have developed allow this map to
be directly related to the form and parameterisation of the quintessence
potential.

Although this work focuses on the actual mass functions of studied
models, that is on the real volume number densities predicted, some
authors have argued that detection of such density should be hindered
by the global geometric effect of dark energy \cite{SoleviEtal04,SoleviEtal05}.
However, their treatment of observed cluster samples with a unique
mass cut-off forgets about the bias-geometry dependence, discussed
by \cite{Kaiser84} and retained as mass-observable relation in dark
energy studies \cite{Mohr04}, that should \textbf{certainly call
for further scrutiny. Such a study is in progress that should test whether a
break of degeneracy is induced by this bias-geometry
dependence and that will be at the core of a companion paper.}

More potentials have attracted attention \cite{Barreiroetal00,Albrecht&Skordis00}
and although they combine our studied exponential and power laws,
they should be confronted using our method. Confirmation could be
sought using parallel semi-analytical methods like the \cite{Jenkinsetal01}
one in the context of quintessence \cite{LinderJenkins03}. Eventually,
the homogeneity and minimal coupling of the field lacks proof in the
highly non-linear regime \cite{MotaVdB00,NunesMota04} and calls for
some wider explorations in the line of \cite{MotaVdB00,NunesMota04,MaccioEtal04}.
This will be the subject of a follow up paper.

\ack

MLeD wishes to thank J.-M. Alimi and the LUTH for their hospitality
and in the initiation of this work, H.Courtois and the CRAL, A. Falvard,
E. Giraud and the GAM-LPTA and J.P. Mimoso and the CFTC, for their
hospitality and support, Tom Theuns for fruitfull discussions
and advice, A. Jenkins for letting me modify his mass function code
\textbf{with my quintessence non-linear collapse program}, and T.
Lehner and D. Steer \textbf{and also R. Bower} for their encouragements, efforts
and many discussions. {\scriptsize }{\scriptsize \par}



\footnotesize
\References
\bibitem[Albrecht \& Skordis 2000]{Albrecht&Skordis00}Albrecht, A., \& Skordis, C., 2000, \emph{Phys. Rev. Lett.}, \textbf{84},
2076 (astro-ph/9908085)
\bibitem[Bahcall \textit{et al.} 1999]{Bahcall99}Bahcall, N., Ostriker, J.P., Perlmutter, S., \& Steinhardt, P.J.,
1999, \emph{Science}, \textbf{284}, 1481 (astro-ph/9906463)
\bibitem[Bardeen \textit{et al.} 1986]{BBKS}Bardeen, J.M., Bond, J.R., Kaiser, N., \& Szalay, A.S., 1986, \emph{Ap.J.},
\textbf{304}, 15 
\bibitem[Barreiro \textit{et al.} 2000]{Barreiroetal00}Barreiro, T., Copeland, E.J., \& Nunes, N.J., 2000, \emph{Phys. Rev. D},
\textbf{61}, 127301 (astro-ph/9910214)
\bibitem[Benabed \& Bernardeau 2001]{BenabedBern01}Benabed, K., \& Bernardeau, F., 2001, \emph{Phys. Rev. D}, \textbf{\noun{64}},
083501 (astro-ph/0104371)
\bibitem[Bennett {\it et al.} 2003]{BennettEtal03}Bennett, C.L. \emph{et al.}, 2003, \emph{Ap.J.S.}, \textbf{148}, 1 (astro-ph/0302207)
\bibitem[Bertschinger 1985]{Bertsching85}Bertschinger, E., 1985, \emph{Ap.J.S.}, \textbf{58}, 39
\bibitem[Biesiada {\it et al.} 2005]{BiesiadaEtal04}Biesiada, M.,
God\l owski, W, \& Szyd\l owski, M., 2005, \emph{Ap.J.}, \textbf{622}, 28 (astro-ph/0403305)
\bibitem[Binetruy 1999]{Binetruy99}Binetruy, P. , 1999, \emph{Phys. Rev. D}, \textbf{60}, 063502 (hep-ph/9810553)
\bibitem[Bond \& Efstathiou 1984]{BondEfstathiou84}\textbf{Bond, J.R., \& Efstathiou, G., 1984,} \textbf{\emph{Ap.J.}}\textbf{,
285, L45 }
\bibitem[Brax \& Martin 1999]{BraxMartin99}Brax, P., \& Martin, J., 1999, \emph{Phys.Lett.}, \textbf{468B}, 40
(astro-ph/9905040)
\bibitem[Brax \& Martin 2000]{BraxMartin00}Brax, P., \& Martin, J., 2000, \emph{Phys. Rev. D}, \textbf{61}, 103502
(astro-ph/9912046)
\bibitem[Brax \textit{et al.} 2000]{Brax00}Brax, P., Martin, J., \& Riazuelo, A., 2000, \emph{Phys. Rev. D}, \textbf{62},
{103505} (astro-ph/0005428)
\bibitem[Carlstrom \textit{et al.} 2002]{Carlstrom02}Carlstrom, J.E., Holder, G.P., \& Reese E.D., 2002, \emph{A.R.A.\&A.}, \textbf{40},
{643} (astro-ph/0208192)
\bibitem[Chen \& Ratra 2004]{ChenRatra04}Chen, G., \& Ratra, B., 2004, \emph{Ap.J.}, \textbf{613}, {L1} (astro-ph/0405636)
\bibitem[Cole \& Lucchin 1995]{ColeLucchin95}Cole,P., \& Lucchin, F. 1995 \emph{Cosmology: The Origin and Evolution
of Cosmic Structure} (Chichester, UK: John Wiley \& Sons)
\bibitem[Daly \& Djorgovski 2004]{DalyDjorgovski04}Daly, R.A., \&
Djorgovski,S.G., 2004, \emph{Ap.J.}, \textbf{612}, 652
(astro-ph/0403664)
\bibitem[Dolag \textit{et al.} 2004]{DolagEtal04}Dolag, K., Bartelmann, M., Perrotta, F., Baccigalupi, C., Moscardini,
L., Meneghetti, M., \& Tormen, G., 2004, \emph{A.\&A.}, \textbf{416}, {853}
\bibitem[Efstathiou  \textit{et al.} 1992]{EfstEtal92}Efstathiou, G., Bond, J. R., \& white S. D. M., 1992, \emph{M.N.R.A.S.},
\textbf{258}, {1}
\bibitem[Efstathiou 2000]{Efst00}Efstathiou, G.,2000, \emph{M.N.R.A.S.}, \textbf{310}, {842} (astro-ph/9904356)
\bibitem[Eke {\it et al.} 1996]{Eke96}Eke, V.R., Cole, S., \& Frenk, C.S., 1996,\emph{M.N.R.A.S.}, \textbf{282}, {263}
(astro-ph/9601088)
\bibitem[Ferreira \& Joyce 1998]{FerreiraJoyce98}Ferreira, P.G., \& Joyce, M., 1998, \emph{Phys. Rev. D}, \textbf{58}, {023503}
(astro-ph/9711102)
\bibitem[Fillmore \& Goldreich 1984]{FG84}Fillmore, J. A. \& Goldreich, P., 1984, \emph{Ap.J.}, \textbf{281}, {1}
\bibitem[Freedman 2000]{Freedman00}Freedman, W.L., 2000, \emph{Physica Scripta}, \textbf{T85}, {37} (astro-ph/9905222)
\bibitem[Gladder \& Yee 2005]{GladderYee05}Gladder, M.D., \& Yee, H., 2005, \emph{Ap.J.S.}, \textbf{157}, {1} (astro-ph/0411075)
\bibitem[Gunn \& Gott 1972]{GG72}Gunn, J. E. \& Gott, J. R., 1972, \emph{Ap.J.}, \textbf{176}, {1}
\bibitem[Hannestad \& M\"ortsell 2004]{HannestadMortsell04}Hannestad, S., \& M\"ortsell, E., 2004, \emph{J.C.A.P.}, \textbf{0409}, {001}
(astro-ph/0407259)
\bibitem[Hannestad 2005]{Hannestad05}Hannestad, S., 2005, \emph{Phys. Rev. D}, \textbf{71}, 103519 (astro-ph/0504017)
\bibitem[Hu 2003]{Hu03}Hu, W., \& Scranton, R., 2004,
\emph{Phys. Rev. D}, \textbf{\noun{70}}, {l23002}
(astro-ph/0408456)
\bibitem[Jenkins \textit{et al} 2001]{Jenkinsetal01}Jenkins, A., Frenk, C.S., White, S.D.M., Colberg, J.M., Evrard, A.E.,
Couchman, H.M.P., \& Yoshida, N., 2001, \emph{M.N.R.A.S.}, \textbf{321},
{372} (astro-ph/0005260)
\bibitem[Kaiser 1984]{Kaiser84}Kaiser, N., 1984, \emph{Ap.J.}, \textbf{284}, {L9} 
\bibitem[Klypin \textit{et al.} 2003]{KlypinEtal03}Klypin, A., Macci\`o, A.V., Mainini, R., \& Bonomento, S.A., 2003,
\emph{Ap.J.}, \textbf{599}, {31} (astro-ph/0303304)
\bibitem[Kuhlen \textit{et al.} 2005]{KuhlenEtal05}Kuhlen, M., Strigari, L.E., Zentner, A.R., Bullock, J.S., \& Primack,
J.R., 2005, \emph{M.N.R.A.S.}, \textbf{357}, {387} (astro-ph/0402210)
\bibitem[Larson 1969]{Larson69}Larson, R. B., 1969, \emph{M.N.R.A.S.}, \textbf{145}, {271}
\bibitem[Linder \& Jenkins 2003]{LinderJenkins03}Linder, E.V., \& Jenkins, A.R., 2003, \emph{M.N.R.A.S.}, \textbf{346}, {573}
(astro-ph/0305286)
\bibitem[Lokas \& Hoffman 2001]{LokasHoffman01}Lokas, E.L., \&
Hoffman, Y. 2001 \emph{Proc. of 3rd Internat. Workshop
on the identification of Dark Matter}, eds. N.J.C. Spooner \& V. Kdryavtsev,
World Scientific, (Singapore) 121 (astro-ph/0011295)
\bibitem[Lokas \textit{et al.} 2004]{LokasEtal04}Lokas, E.L., Bode, P., \& Hoffman, Y., 2004, \emph{M.N.R.A.S.}, \textbf{349},
{595} (astro-ph/0309485)
\bibitem[Ma \textit{et al.} 1999]{Maetal99}Ma, C.-P., Caldwell, R.R., Bode, P., \& Wang, L., 1999, \emph{Ap.J.}, \textbf{521},
{L1} (astro-ph/9906174)
\bibitem[Macci\`o \textit{et al.} 2004]{MaccioEtal04}Macci\`o, A.V., Quercellini, C., Mainini, R., Amendola, L., \& Bonomento,
S.A., 2004, \emph{Phys. Rev. D}, \textbf{69}, {123516} (astro-ph/0309671)
\bibitem[Mainini \textit{et al.} 2003a]{MaininiEtal03a}Mainini, R.,
Macci\`o, A.V., \& Bonomento, S.A., 2003, \emph{New Astron.},
\textbf{8}, {173} (astro-ph/0207581)
\bibitem[Mainini \textit{et al.} 2003b]{MaininiEtal03b}Mainini, R., Macci\`o, A.V., Bonomento, S.A., \& Klypin, A., 2003,
\emph{Ap.J.}, \textbf{599}, {24} (astro-ph/0303303)
\bibitem[Majumdar \& Mohr 2003]{Majumdar03}Majumdar, S., \& Mohr, J.J., 2004, \emph{Ap.J.}, \textbf{\noun{613}}, {41} (astro-ph/0305341)
\bibitem[Mohr 2004]{Mohr04}Mohr, J.J. 2004 \emph{proc. of NOAO: Observing Dark Energy} (astro-ph/0408484)
\bibitem[Mota \& van de Bruck 2004]{MotaVdB00}Mota, D.F., \& van de Bruck, C., 2004, \emph{A.\&A.}, \textbf{421}, {71}
(astro-ph/0401504)
\bibitem[Nesseris \& Perivolaropoulos 2004]{NesserisPeriv04}Nesseris,
S., \& Perivolaropoulos, L., 2004, \emph{Phys. Rev. D}, \textbf{70},
{043531} (astro-ph/0401556)
\bibitem[Nunes \& Mota 2004]{NunesMota04}Nunes, N.J., \& Mota, D.F.
2004 Structure Formation in Inhomogeneous Dark Energy Models
\emph{Preprint, for M.N.R.A.S.}, astro-ph/0409481
\bibitem[Padmanabhan 2003]{Padman03}Padmanabhan, T., 2003,
\emph{Phys. Rept.}, \textbf{380}, 235 (hep-th/0212290)
\bibitem[Page \textit{et al.} 2003]{PageEtal03}Page, L. \emph{\
et al.}, 2003, \emph{Ap.J.S.}, \textbf{148}, {233} (astro-ph/0302220)
\bibitem[Peebles 1980]{Peebles}Peebles, P.J.E., 1980, \emph{The Large Scale Structure of the Universe}
(Princeton: Princeton University Press)
\bibitem[Penston 1969]{Penston69}Penston, M. V., 1969, \emph{M.N.R.A.S.}, \textbf{144}, {425}
\bibitem[Perlmutter \textit{et al.} 1998]{Perlmut98}Perlmutter, S. \emph{\
et al.}, 1999, \emph{Ap.J.}, \textbf{517}, {565} (astro-ph/9812133)
\bibitem[Press \& Schechter 1974]{PS}Press, W.H., \& Schechter, P., 1974, \emph{Ap.J.}, \textbf{187}, {425}
\bibitem[Ratra \& Peebles 1988]{RP}Ratra, B., \& Peebles, P.J.E., 1988, \emph{Phys. Rev. D}, \textbf{37}, {3406}
\bibitem[Ruhl \textit{et al.} 2004]{Ruhl04}Rhul , J.E. \emph{et al.},
2004, \emph{S.P.I.E.}, \textbf{5498}, {11} (astro-ph/0411122)
\bibitem[Riess \textit{et al.} 1998]{Riess98}Riess, A.G. \emph{et al.}, 1998, \emph{A.J.}, \textbf{116}, {1009} (astro-ph/9805201)
\bibitem[Riess\,\textit{et\,al.}\,2004]{Riess04}Riess, A.G. \emph{et al.}, 2004, \emph{Ap.J.}, \textbf{607}, {665} (astro-ph/0402512)
\bibitem[Sahni 2004]{Sahni04}Sahni, V. 2004 \emph{2nd Aegean Summer School on the Early Universe} (astro-ph/0403324)
\bibitem[Solevi \textit{et al.} 2004]{SoleviEtal04}Solevi, P., Mainini,
R., \& Bonomento, S.A. 2004 The Nature of Dark Energy from deep
Cluster Abundance \emph{Preprint, for Ap.J.}, astro-ph/0412054
\bibitem[Solevi \textit{et al.} 2005]{SoleviEtal05}Solevi, P., Mainini, R., \& Bonomento, S.A., Macci\`o, A.V., Klypin,
A., \& Gottl\"ober, S. 2005 Tracing the Nature of Dark Energy with
Galaxy Distribution \emph{Preprint for M.N.R.A.S.} astro-ph/0504124
\bibitem[Steinhardt \textit{et al.} 1999]{Steinhardtetal99}Steinhardt, P.J., Wang, L., \& Zlatev, I., 1999, \emph{Phys. Rev. D}, \textbf{59},
{123504} (astro-ph/9812313)
\bibitem[Sugiyama 1995]{Sugiyama95}Sugiyama, N., 1995, \emph{Ap.J.S.}, \textbf{100}, {281} (astro-ph/9412025)
\bibitem[Wang \textit{et al.} 2000]{Wangetal00}Wang, L., Caldwell, R.R., Ostriker, J.P., \& Steinhardt, P.J., 2000,
\emph{Ap.J.}, \textbf{530}, {17} (astro-ph/9901388)
\bibitem[Wang \& Mukherjee 2004]{WangMukh04}Wang, Y., \& Mukherjee, P., 2004, \emph{Ap.J.}, \textbf{379}, {652} (astro-ph/0312192)

\endrefs

\end{document}